\documentclass[twoside,twocolumn,9pt]{article}
\usepackage{extsizes}
\usepackage[super,sort&compress,comma]{natbib} 
\usepackage[version=3]{mhchem}
\usepackage[left=1.5cm, right=1.5cm, top=1.785cm, bottom=2.0cm]{geometry}
\usepackage{balance}
\usepackage{mathptmx}
\usepackage{amssymb}
\usepackage{sectsty}
\usepackage{graphicx} 
\usepackage{lastpage}
\usepackage[format=plain,justification=justified,singlelinecheck=false,font={stretch=1.125,small,sf},labelfont=bf,labelsep=space]{caption}
\usepackage{float}
\usepackage{fancyhdr}
\usepackage{fnpos}
\usepackage[english]{babel}
\addto{\captionsenglish}{%
  
}
\usepackage{array}
\usepackage{droidsans}
\usepackage{charter}
\usepackage[T1]{fontenc}
\usepackage[usenames,dvipsnames]{xcolor}
\usepackage{setspace}
\usepackage[compact]{titlesec}
\usepackage{hyperref}

\usepackage{epstopdf}

\definecolor{cream}{RGB}{222,217,201}

\begin{document}

\nocite{rsc-control}

\pagestyle{fancy}
\thispagestyle{plain}
\fancypagestyle{plain}{
\renewcommand{\headrulewidth}{0pt}
}

\makeFNbottom
\makeatletter
\renewcommand\LARGE{\@setfontsize\LARGE{15pt}{17}}
\renewcommand\Large{\@setfontsize\Large{12pt}{14}}
\renewcommand\large{\@setfontsize\large{10pt}{12}}
\renewcommand\footnotesize{\@setfontsize\footnotesize{7pt}{10}}
\makeatother

\renewcommand{\thefootnote}{\fnsymbol{footnote}}
\renewcommand\footnoterule{\vspace*{1pt}%
\color{cream}\hrule width 3.5in height 0.4pt \color{black}\vspace*{5pt}} 
\setcounter{secnumdepth}{5}

\makeatletter 
\renewcommand\@biblabel[1]{#1}            
\renewcommand\@makefntext[1]%
{\noindent\makebox[0pt][r]{\@thefnmark\,}#1}
\makeatother 
\renewcommand{\figurename}{\small{Fig.}~}
\sectionfont{\sffamily\Large}
\subsectionfont{\normalsize}
\subsubsectionfont{\bf}
\setstretch{1.125} 
\setlength{\skip\footins}{0.8cm}
\setlength{\footnotesep}{0.25cm}
\setlength{\jot}{10pt}
\titlespacing*{\section}{0pt}{4pt}{4pt}
\titlespacing*{\subsection}{0pt}{15pt}{1pt}

\fancyfoot{}
\fancyfoot[LO,RE]{\vspace{-7.1pt}\includegraphics[height=9pt]{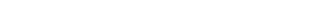}}
\fancyfoot[CO]{\vspace{-7.1pt}\hspace{13.2cm}\includegraphics{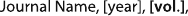}}
\fancyfoot[CE]{\vspace{-7.2pt}\hspace{-14.2cm}\includegraphics{head_foot/RF}}
\fancyfoot[RO]{\footnotesize{\sffamily{1--\pageref{LastPage} ~\textbar  \hspace{2pt}\thepage}}}
\fancyfoot[LE]{\footnotesize{\sffamily{\thepage~\textbar\hspace{3.45cm} 1--\pageref{LastPage}}}}
\fancyhead{}
\renewcommand{\headrulewidth}{0pt} 
\renewcommand{\footrulewidth}{0pt}
\setlength{\arrayrulewidth}{1pt}
\setlength{\columnsep}{6.5mm}
\setlength\bibsep{1pt}

\makeatletter 
\newlength{\figrulesep} 
\setlength{\figrulesep}{0.5\textfloatsep} 

\newcommand{\topfigrule}{\vspace*{-1pt}%
\noindent{\color{cream}\rule[-\figrulesep]{\columnwidth}{1.5pt}} }

\newcommand{\botfigrule}{\vspace*{-2pt}%
\noindent{\color{cream}\rule[\figrulesep]{\columnwidth}{1.5pt}} }

\newcommand{\dblfigrule}{\vspace*{-1pt}%
\noindent{\color{cream}\rule[-\figrulesep]{\textwidth}{1.5pt}} }

\makeatother

\newcommand{\apjs}{ApJS}
\newcommand{\apj}{ApJ}
\newcommand{\apjl}{ApJ}
\newcommand{\mnras}{MNRAS}
\newcommand{\aap}{A\&A}
\newcommand{\aj}{AJ}
\newcommand{\nat}{Nature}
\newcommand{\bain}{Bull.~Astron.~Inst.~Netherlands} 
\newcommand{\araa}{ARA\&A}
\newcommand{\icarus}{Icarus}
\newcommand{\ssr}{Space~Sci.~Rev.}
\newcommand{\pasp}{PASP}
\newcommand{\jqsrt}{JQRST}

\twocolumn[
  \begin{@twocolumnfalse}
{\includegraphics[height=30pt]{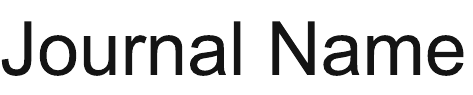}\hfill\raisebox{0pt}[0pt][0pt]{\includegraphics[height=55pt]{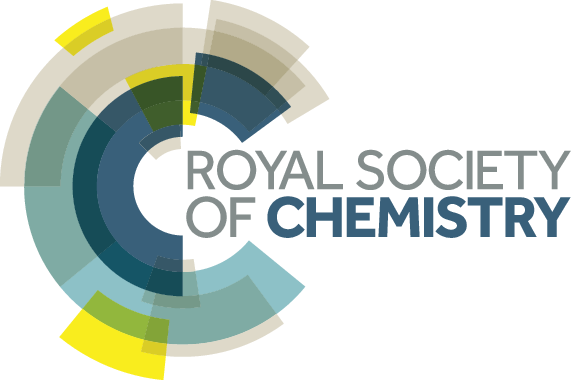}}\\[1ex]
\includegraphics[width=18.5cm]{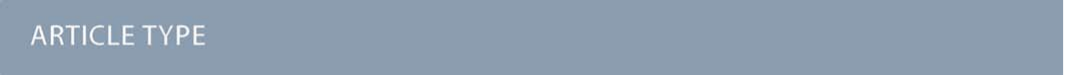}}\par
\vspace{1em}
\sffamily
\begin{tabular}{m{4.5cm} p{13.5cm} }

\includegraphics{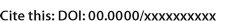} & \noindent\LARGE{\textbf{The Chemical Inventory of the Inner Regions of Planet-forming Disks  – The JWST/MINDS Program$^\dag$}} \\
\vspace{0.3cm} & \vspace{0.3cm} \\

 & \noindent\large{Inga Kamp,$^{\ast}$\textit{$^{a}$} Thomas Henning,\textit{$^{b}$} Aditya M.\ Arabhavi\textit{$^{a}$}, Giulio Bettoni\textit{$^{g}$}, Valentin Christiaens\textit{$^{d}$}, Danny Gasman\textit{$^{e}$}, Sierra L.\ Grant\textit{$^{g}$}, Maria Morales-Calder\'on\textit{$^{f}$}, Beno\^{i}t Tabone\textit{$^{c}$}, Alain Abergel\textit{$^{c}$}, Olivier Absil\textit{$^{d}$}, Ioannis Argyriou\textit{$^{e}$}, David Barrado\textit{$^{f}$}, Anthony Boccaletti\textit{$^{h}$}, Jeroen Bouwman\textit{$^{b}$}, Alessio Caratti o Garatti\textit{$^{i,j}$}, Ewine F.\ van Dishoeck\textit{$^{g,k}$}, Vincent Geers\textit{$^{l}$}, Adrian M.\ Glauser\textit{$^{m}$}, Manuel G\"udel\textit{$^{b,m,n}$}, Rodrigo Guadarrama\textit{$^{n}$}, Hyerin Jang\textit{$^{o}$}, Jayatee Kanwar\textit{$^{a,p}$}, Pierre-Olivier Lagage\textit{$^{q}$}, Fred Lahuis\textit{$^{r}$}, Michael Mueller\textit{$^{a}$}, Cyrine Nehm\'e\textit{$^{s}$}, G\"oran Olofsson\textit{$^{t}$}, Eric Pantin\textit{$^{u}$}, Nicole Pawellek\textit{$^{n,w}$}, Giulia Perotti\textit{$^{b}$}, Tom P.\ Ray\textit{$^{j}$}, Donna Rodgers-Lee\textit{$^{j}$}, Matthias Samland\textit{$^{b}$}, Silvia Scheithauer\textit{$^{b}$}, J\"urgen Schreiber\textit{$^{b}$}, Kamber Schwarz\textit{$^{b}$}, Milou Temmink\textit{$^{k}$}, Bart Vandenbussche\textit{$^{e}$}, Marissa Vlasblom\textit{$^{k}$}, Christoffel Waelkens\textit{$^{e}$}, L.\ B.\ F.\ M.\ Waters\textit{$^{o,v}$}, Gillian Wright\textit{$^{l}$}} 
 \\

\includegraphics{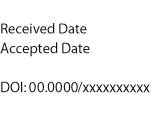} & \noindent\normalsize{The understanding of planet formation has changed recently, embracing the new idea of pebble accretion. This means that the influx of pebbles from the outer regions of planet-forming disks to their inner zones could determine the composition of planets and their atmospheres.
The solid and molecular components delivered to the planet-forming region can be best characterized by mid-infrared spectroscopy. With Spitzer low-resolution ($R\!=\!100$, 600) spectroscopy, this approach was limited to the detection of abundant molecules such as H$_2$O, C$_2$H$_2$, HCN and CO$_2$. This contribution will present first results of the MINDS (MIRI mid-IR Disk Survey, PI: Th.\ Henning) project. Due do the sensitivity and spectral resolution ($R\!\sim\!1500-3500$) provided by JWST we now have a unique tool to obtain the full inventory of chemistry in the inner disks of solar-types stars and brown dwarfs, including also less abundant hydrocarbons and isotopologues. The Integral Field Unit (IFU) capabilities enable at the same time spatial studies of the continuum and line emission in extended sources such as debris disks, the flying saucer and also the search for mid-IR signatures of forming planets in systems such as PDS\,70. These JWST observations are complementary to ALMA and NOEMA observations of the outer disk chemistry; together these datasets provide an integral view of the processes occurring during the planet formation phase.} \\

\end{tabular}

 \end{@twocolumnfalse} \vspace{0.6cm}

  ]

\renewcommand*\rmdefault{bch}\normalfont\upshape
\rmfamily
\section*{}
\vspace{-1cm}


\footnotetext{\textit{$^{a}$~Kapteyn Astronomical Institute, University of Groningen, PO BOX 800, 9700 AV Groningen, The Netherlands; E-mail: kamp@astro.rug.nl}}

\footnotetext{\textit{$^{b}$~Max-Planck-Institut f\"{u}r Astronomie (MPIA), K\"{o}nigstuhl 17, 69117 Heidelberg, Germany. }}

\footnotetext{\dag~The MIRI mid-INfrared Disk Survey (MINDS) is part of the JWST MIRI GTO time and led by Thomas Henning and Inga Kamp.}



\section{Introduction}

Much of the exoplanet population studied so far resides inside 10~au from their host star. The composition of these planets should carry traces of the chemical composition of the inner regions of planet-forming disks. This can manifest itself in the bulk C/O ratio of gas giant planet atmospheres \cite{Mordasini2016, Cridland2020, Schneider2021, Molliere2022}, but also affect the bulk interior composition of terrestrial planets (e.g.\ the sulphur content) \cite{Jorge2022} and the delivery of water to them \cite{Izidoro2013, Ciesla2015}. 

\begin{figure}[t]
\centering
  \includegraphics[width=8cm]{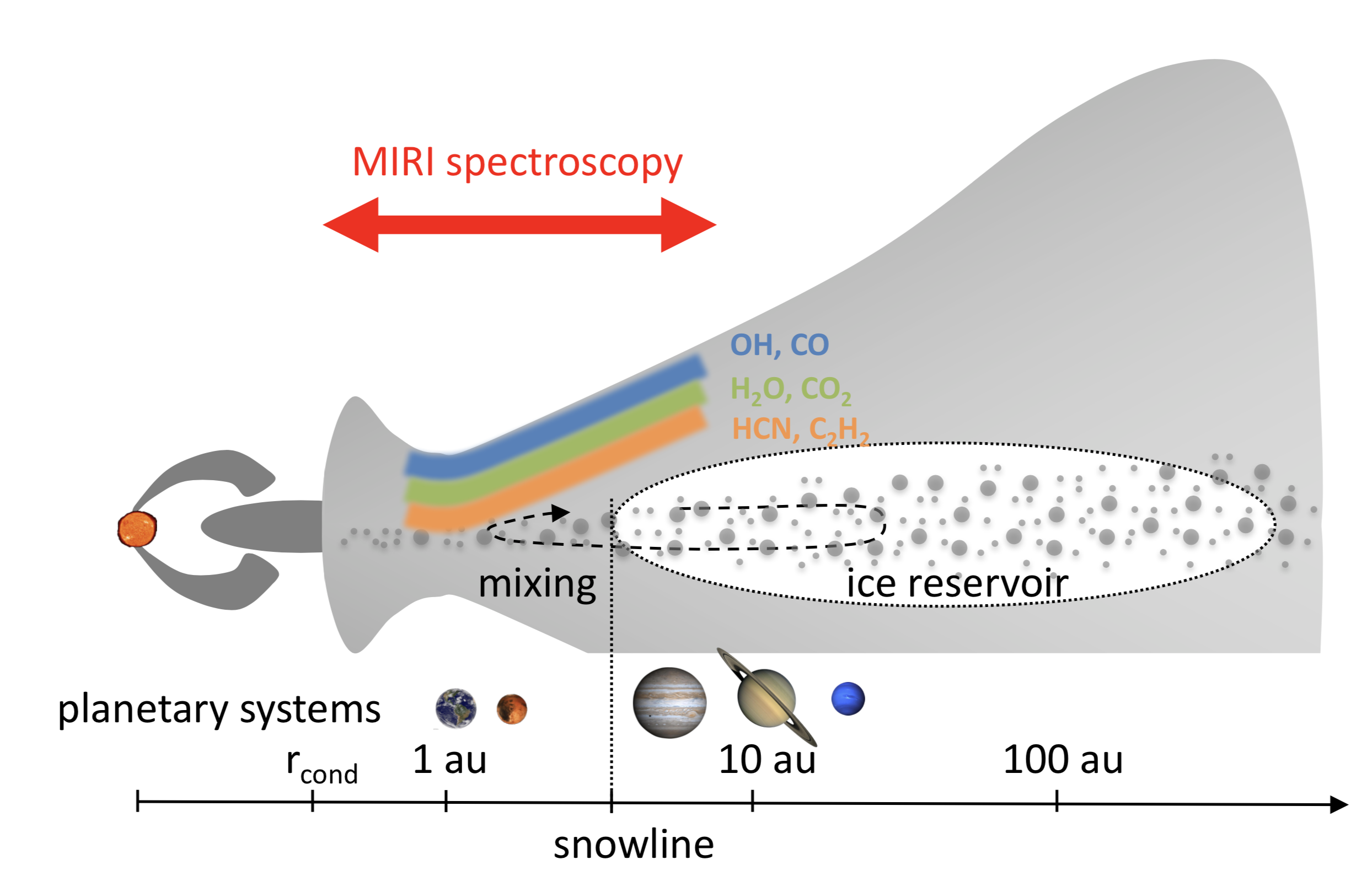}
  \caption{Sketch of a typical disk around a T Tauri star. Thermo-chemical disk models suggest a layered molecular structure with OH and CO at the top, followed by H$_2$O and CO$_2$. HCN and C$_2$H$_2$ peak in abundance deeper in the disk. Theory and observations suggest the possibility of radial transport of icy grains which can affect the inner disk gas composition. The MIRI instrument on board the James Webb Space Telescope probes the inner ($\lesssim\,10$~au) warm disk surface.}
  \label{fig:disk-sketch}
\end{figure}

Planet-forming disks are expected to be layered in their chemical content, with the surface layers being ionized/atomic and deeper layers being molecular \cite{Bergin2007, Henning2013}. More recently, thermo-chemical models also showed that the spatial distribution of molecules comes in layers, with OH being closest to the surface and CO, H$_2$O, CO$_2$, HCN and C$_2$H$_2$ residing ever closer to the midplane \cite{Woitke2018} (Fig.~\ref{fig:disk-sketch}). In addition, theory and observations have shown that dust grains can radially migrate in the disk if they reach sizes that allow them to dynamically de-couple from the gas \cite{Birnstiel2016,Andrews2020}. As a consequence, volatile ices carried along and sublimating at the respective iceline locations could alter the elemental composition of the gas in the disk \cite{Krijt2020}. The inner disk regions are generally highly optically thick (unless some processes have removed material and carved gaps/holes\cite{Espaillat2014}), preventing us from probing down to the midplane. However turbulent mixing timescales are less than $10\,000$~yr \cite{Semenov2011}, thus ensuring that any change of element abundance due to radial transport in the midplane should quickly spread vertically in the disk.

The mid-infrared wavelength range is key in probing the inner $\sim\!10$~au of planet-forming disks. The gas in there is typically warm (several 100-1000~K) and we expect to see a mix of highly excited rotational lines and ro-vibrational ones. The Spitzer Space Telescope has frequently detected abundant molecules such as water, OH, CO$_2$, HCN and C$_2$H$_2$ in disks around Solar-type stars, and less frequently around Herbig Ae stars \cite{Carr2008, Pontoppidan2010, Salyk2011, Najita2013}. Still, due to the low spectral resolution of $R\!\sim\!600$, Spitzer saw only the tip of the iceberg; less abundant molecules (minor species, e.g.\ CH$_4$ and NH$_3$ with features $<10~\mu$m, and other hydrocarbons such as C$_4$H$_2$) remained undetected. Yet these minor species play a major role in the synthesis pathways of complex organic molecules and are key to understanding the types of chemistry occurring while planet formation is in full steam (e.g.\ C- or O-chemistry, combustion chemistry). Also, the quantitative interpretation of the molecular bands in the Spitzer data was strongly hampered by line blending of the various molecules. The recently launched James Webb Space Telescope (JWST) has the spectral resolution ($R\!\sim\!1500-3500$) and the sensitivity to disentangle major species blends and to carry out a deep search for minor species.

In the following, we will present some first results on the richness of the inner disks of T Tauri and low-mass stars from Guaranteed Time Observations (GTO) with the MIRI instrument on board of the James Webb Space Telescope.

\begin{figure}[t]
\centering
  \vspace*{-5mm}
  \includegraphics[width=9cm]{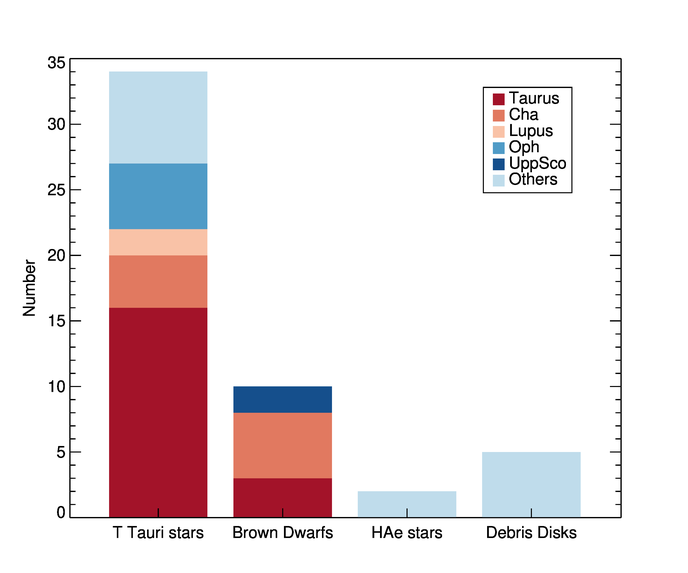}
  \caption{Distribution of the 51 MINDS sources across planet-forming disk type and star-forming region.}
  \label{fig:MINDS-sample}
\end{figure}

\section{The MINDS survey}

The MIRI mid-INfrared Disk Survey (MINDS) is part of the JWST MIRI GTO time and led by Thomas Henning and Inga Kamp (PID: 1282). We use the medium resolution spectrograph (MRS) of the Mid-InfraRed Instrument MIRI \cite{Rieke2015, Wright2015}, which covers simultaneously the wavelength range 4.9 to 28.6\,$\mu$m with a spectral resolution of $R\!\sim\!1500-3500$. 

\subsection{The sample}

We selected our sample to contain young gas-rich class\,{\sc ii} disks in nearby star-forming regions (Herbig Ae stars, T Tauri stars and very low mass stars/brown dwarfs). Brown dwarfs are sub-stellar objects that are not massive enough to sustain stable hydrogen fusion. The boundary between very low mass stars (VLMS) and brown dwarfs (BDs) is not well defined; some authors use color criteria \cite{Alvesdeoliveira2013,Tang2018} or spectral type (later than M6 for young BDs \cite{Luhman2012}). In addition, we selected a handful of young gas-rich debris disks, where ALMA has established the presence of CO gas. We mostly selected targets in the Taurus, Ophiuchus, Chamaeleon and Lupus star-forming regions largely based on the presence of molecular emission known from previous Spitzer observations and using ground-based CO ro-vibrational line profiles to make sure we look at disk emission. Figure~\ref{fig:MINDS-sample} shows the distribution of our targets across the various star-forming regions and object types; our main focus is on the Taurus star-forming region. We also included in our sample PDS\,70, a young planet-forming disk, which is known to host two gas giant protoplanets \cite{Keppler2018, Haffert2019}, and TW\,Hya, one of the closest gas-rich planet-forming disks ($d\!=\!60$~pc). Both systems have the potential to reveal intriguing mid-IR signatures related to forming planets.

\subsection{The MIRI/MRS survey}

The total observing time of our survey is $\sim\,120$~h. We base our exposure time estimates on previous mid-IR Spitzer fluxes and the goal to obtain a S/N of $300-500$ to maximize detection rates of minor species \cite{Bosman2017}. Due to the instrument sensitivity being a strong function of wavelengths and the Spectral Energy Distribution (SED) of our targets being non-flat, we will mostly achieve this S/N at shorter wavelengths. Since all our targets are relatively bright sources with well known coordinates and proper motions (Gaia DR3 \cite{Gaia2022}), we omit a dedicated target acquisition (TA) with the exception of the brown dwarfs where TA is necessary. For the spatially extended sources in our sample we schedule a dedicated background observation just before the science exposure with the same exposure time. For the science exposures, we use a 4-point dither strategy.

\subsection{Data reduction}

Our data reduction flow leverages both the JWST pipeline (v1.8.4) and routines from the VIP package \cite{GomezGonzalez2017}. From the JWST pipeline, we use (i) the class \texttt{Detector1} to process uncalibrated raw data files using CRDS context \texttt{jwst\_1019.pmap} and default parameters; (ii) \texttt{Spec2} with default parameters, but skipping background subtraction, and using dedicated reference files for photometric and fringe flat calibrations as in Gasman et al. (subm., arXiv:2212.03596); (iii) \texttt{Spec3} with default parameters, apart from the \texttt{master\textunderscore background} and \texttt{outlier\textunderscore detection} steps which are turned off. 
We leverage the four-point dither strategy to obtain a first guess of the background signal as the minimum of each quadruplet of dithered images (on detector), then refined this estimate using a median-filter, 
which both smoothed the background map and removed residual star signals. This estimated background map is subtracted just before \texttt{Spec3}. Bad pixel correction is performed both before background estimation and after \texttt{Spec3} using dedicated VIP routines. Spectrum extraction is performed manually through aperture photometry in 2.5 $\lambda/D$ apertures in the final spectral frames, after identifying manually the star centroid. We use correction factors to compensate for the missing flux in the apertures (as in Argyriou et al., in prep.), and filter out spikes affecting individual spaxels when integrating the signal in each aperture.

\section{Small dust grains and gas in the inner disk}

We have little direct imaging information on the detailed gas and dust structure of the inner 10~au of disks around low-mass stars ($M_\ast\!\lesssim\!1$~M$_\odot$). Much of what we know originates from the kinematic analysis of line profiles such as the CO ro-vibrational lines \cite{Brittain2009, Banzatti2017}, spectroscopy of dust emission features \cite{Olofsson2009} or interferometry \cite{Anthonioz2015}. The Spitzer Space Telescope has studied samples of T\,Tauri disks down to the brown dwarfs regime both for their molecular content \cite{Pascucci2009, Pontoppidan2010, Carr2011, Salyk2011b, Najita2013, Pascucci2013} and their dust mineralogy and grain sizes \cite{Olofsson2009, Pascucci2009, Olofsson2010}. The inner disk surfaces are generally populated by micron-sized grains and disk models suggest that the 10 and 20~$\mu$m silicate feature and the molecular lines originate from a similar radial region.

\begin{figure}[t]
\centering
  \includegraphics[width=8cm]{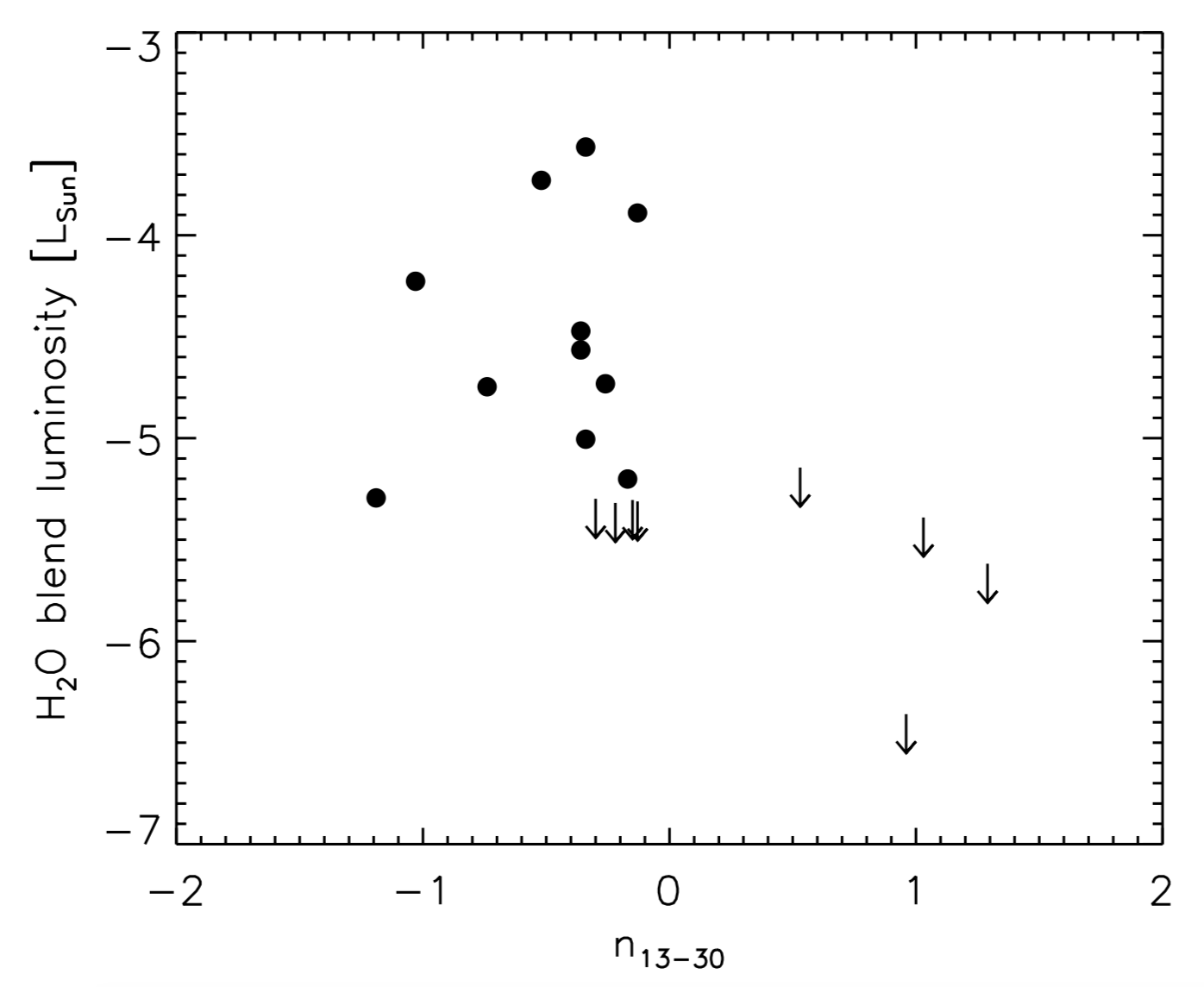}
  \caption{Water luminosity in logarithmic units versus SED slope between 13 and 30\,$\mu$m for MINDS T\,Tauri disks that have high resolution Spitzer spectra ($R\!\sim\!600$). Arrows are upper limits. The higher spectral resolution and sensitivity of JWST MIRI/MRS allows to detect much fainter water fluxes in the MINDS program.} 
  \label{fig:water-vs-SEDslope}
\end{figure}

Figure~\ref{fig:water-vs-SEDslope} shows the Spitzer water luminosity\cite{Banzatti2020} vs SED slope $n_{13-30}$ (MSc thesis Christian Lemmens) for a subset of our 19 T\,Tauri MINDS targets. Like previous work \cite{Banzatti2020}, we see no obvious correlation across our sample; the only interesting point is that disks with large cavities ($n_{13-30}\!>\!0.5$) have no water detection. It remains to be seen whether this was a sensitivity limit. Given the typical MINDS integration times, we measure noise levels of $\sim\,2-12$~mJy around 17\,$\mu$m \citep{Grant2022}, corresponding to line luminosities of $5\cdot 10^{-8}-3\cdot 10^{-7}$~L$_\odot$. Theoretical models predict for example water lines to be weaker in disks with inner cavities \cite{Antonellini2015, Anderson2021}.

Overall, we have not yet fully understood where the large diversity of molecular line flux strength originates from; in several cases, line fluxes themselves correlate well, e.g.\ water and CO ro-vib \cite{Salyk2011b}, water and HCN \cite{Najita2013}. It is unclear whether this is because these molecules are co-spatial radially and vertically. Disk models can often reproduce such correlations \cite{Antonellini2023}. Some molecules are not emitting in Local Thermodynamic Equilibrium (LTE); OH is for example non-thermally excited through water photodissociation \cite{Tabone2021}; CO can be pumped by fluorescence or IR radiation \cite{Brittain2009, Thi2013}; HCN can be pumped by IR radiation \cite{Bruderer2015}. Also, the C$_2$H$_2$ fluxes do not correlate with water fluxes \cite{Salyk2011b}. Even in T\,Tauri disks, the strengths of the C$_2$H$_2$ ro-vibrational band is difficult to reproduce with simple gas phase chemistry \cite{Willacy2009, Greenwood2019b}. Higher abundances and column densities of this molecule occur in disk surfaces if the H/H$_2$ transition resides closer to the surface and temperatures are higher \cite{Najita2011}; this can be achieved by e.g.\ considering only small grains instead of a wide grain size distribution up to mm-sizes and settling \cite{Agundez2008, Walsh2015}.

There is still much to learn about the chemical composition of the inner 10~au and the Spitzer spectral resolution was a major limiting factor also in the analysis of the spectra. The blending of lines from different species was difficult to correct without large assumptions and the retrieval suffered from large degeneracies in column densities and temperatures. However, for understanding planetary diversity we need to obtain a quantitative inventory of the inner disk molecular reservoir. The retrieval process will remain the most challenging part of the analysis. Hence, as in times of Spitzer data, we will start simple using 0D slab models and discuss towards the end pathways forwards.

\section{Slab model retrieval}

This is the most simple retrieval method that can be applied to molecular emission spectra. It has been widely used on Spitzer data \cite{Salyk2011}\footnote[3]{slabspec - 10.5281/zenodo.4037306} and using radial temperature gradients (1D) also to near-IR ground based spectroscopy \cite{Brittain2009}. While the method is fast and powerful, it has not yet been assessed how the quantities temperature, column density and emitting area that we retrieve compare to the actual underlying disk properties. The subsequent paragraphs illustrate this and discuss remaining issues and future challenges for retrieval methods.

\subsection{Slab model fluxes}

\begin{figure}[h]
\centering
  \includegraphics[width=3cm]{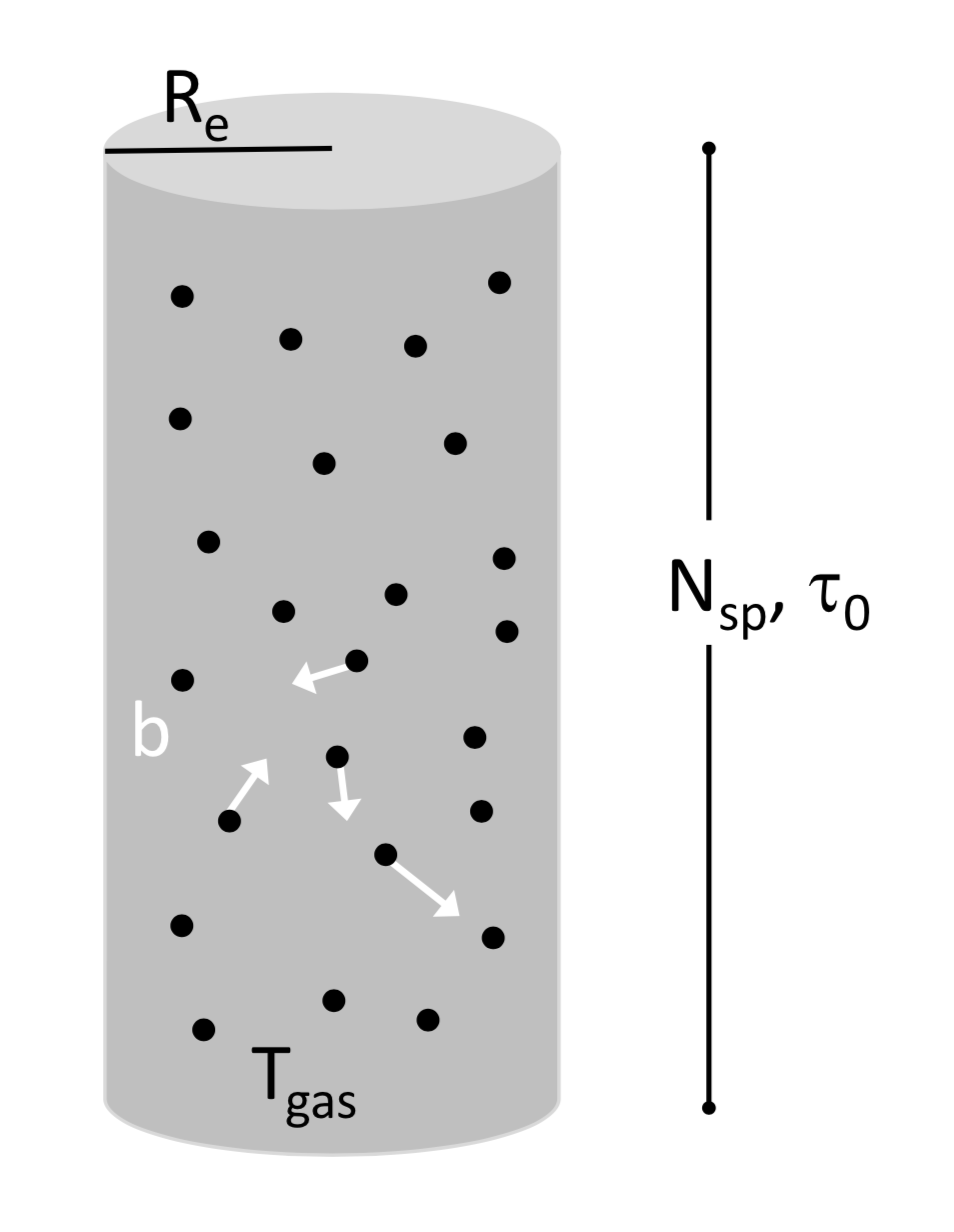}
  \caption{Schematic of a 0D slab model.}
  \label{fig:slab-sketch}
\end{figure}

The most simple 0D slab models calculate the line flux $F_0$ at the center of the line $\lambda_0$ for a selected number of line transitions of a single molecule (species sp). The slab of gas has a constant temperature $T_{\rm gas}$ and density $n_{\rm \langle H \rangle}$\footnote[4]{This is the total hydrogen number density in the gas, so $n({\rm H})+2n({\rm H_2})$.}, and variable length, hence column density $N_{\rm sp}$ (Fig.~\ref{fig:slab-sketch}). Within the slab, the abundance of the species $\epsilon_{\rm sp}$ as well as the potential collision partners (electrons, protons, H, o-/p-H$_2$, He) are free parameters, but constants. The velocity at which gas particles move $b$ is the sum of their thermal velocity $v_{\rm th}$ and a turbulent component $v_{\rm turb}$
\begin{equation}
b=\sqrt{v_{\rm th}^2 + v_{\rm turb}^2} \,\,\,.
\end{equation}

The data characterizing a single line transition of a species sp are the lower and upper energy levels $E_{\rm l}$, $E_{\rm u}$, the respective statistical weights $g_{\rm l}$, $g_{\rm u}$ and the Einstein A coefficient $A_{ul}$. The level populations are denoted as $b_{\rm l}$ and $b_{\rm u}$ and a superscript $^*$ denotes Local Thermodynamic Equilibrium (LTE). The total line intensity emerging $I_0$ is then calculated either using a simple RT solution or from escape probability using
\begin{equation}
I_0 = \frac{h \nu A_{\rm ul}}{4\pi}N_{\rm \langle H \rangle} \epsilon_{sp} b_{\rm u} \beta_{\rm esc}(\tau_0) \,\,\, .
\end{equation}
The Planck constant $h$ and all other quantities are usually used in cgs units. The escape probability $\beta_{\rm esc}$ treatment depends on the geometry assumed \cite{Woitke2009} and the line center optical depth $\tau_0=\tau(\lambda_0)$ 
\begin{equation}
\tau_0 = \frac{A_{\rm ul} \lambda_0^3}{8 \pi b} N_{\rm sp} (b_{\rm l} \frac{g_{\rm u}}{g_{\rm l}} - b_{\rm u})\,\,\, .
\end{equation}
For optically thin lines, the intensity scales directly with the column density of the species. However, as the line become optically thick in the center, the line broadening ensures that the flux keeps growing, albeit not linear anymore. The line photons do then not escape at the line center, but in the wings. In the most simple case, we assume a Gaussian line profile and high optical depth leads to a flat topped quasi rectangular profile.  

The line intensity is converted into a flux assuming a specific emitting area $\pi R_{\rm e}^2$ and the distance $d$ to the source. For a specific turbulence, we can pre-calculate a grid of slab models for a range of temperatures and column densities and this can be re-used for all spectra. 

The level populations can be calculated either from LTE or from Statistical Equilibrium (SE). In the latter case, we need collisional cross sections for all transitions and potential collision partners. The availability of such data is often the limiting factor. Hence, we revert back to LTE for many of the molecules in the mid-IR spectral region. Deviations from LTE will occur if the gas densities are much smaller than the critical density of the line and/or if the molecules are subject to strong radiation fields (e.g.\ UV pumping for CO ro-vibrational lines \cite{Thi2013}, IR pumping by dust thermal emission for HCN \cite{Bruderer2015}) and/or if the excitation occurs through chemical pathways (e.g.\ oxygen through photodissociation of OH \cite{Stoerzer2000}, OH through photodissociation of water \cite{Tabone2021}). It is interesting to note that radiative pumping can drive molecules also into LTE, but the excitation temperature corresponds then to that of the radiation field and not to the gas temperature. Calculating level populations $b_{\rm u}$ wrt.\ the total population of the species in LTE requires knowing the partition functions $Z(T,P)$
\begin{equation}
b_{\rm u} = \frac{n_{\rm sp, u}}{n_{\rm sp, tot}} = \frac{g_{\rm u}}{Z(T,P)} \exp^{-E_{\rm u}/kT}
\end{equation}
with $E_{\rm u}$ the upper levels energy and $k$ the Boltzmann constant. For many molecules these are provided in the form of tables covering a large range of temperature and pressures. In the mid-IR we have to drop the simplified approach of using only a restricted number of levels to calculate the partition functions, an approach often used when dealing with rotational lines at submm wavelength and very cold temperatures ($\lesssim\,100$~K).

We take the molecular data and partition functions mostly from three databases: The Leiden Atomic and Molecular Database (LAMDA \cite{Schoier2005, vanderTak2020}), the HITRAN database \cite{Gordon2022}, and the GEISA database \cite{Delahaye2021}. The GEISA database lacks Einstein A coefficients and those are provided in selected cases by Agnes Perrin (private communication). We also note that several molecules have data only for a single or few ro-vibrational bands. Since the typical temperatures in the inner disks are a few 100 up to 1000~K, our analysis is severely hampered by this. We will come back to it later.

\subsection{Slab model spectra and fitting procedure}

The slab models outlined above provide total line fluxes in a certain wavelength range per molecule. For comparison with actual data, we need to convolve these fluxes with the appropriate spectral resolution of the instrument, in this case $R\!\sim\!3000$ (JWST MIRI/MRS) using a very high spectral resolution wavelength grid. We then sample the spectra using the wavelength points of the observed spectra, thus reproducing the same oversampling as the instrument, and also facilitating the calculation of a formal (reduced) $\chi^2$ in the fitting procedure. 

Since the molecular emission of many molecules overlaps in the mid-IR spectra and the slab models can only deal with one molecule at a time, we define very restricted wavelength windows for the $\chi^2$-minimalization in which we can be sure that only the molecule under investigation contributes. Often these are regions around the strong Q-branches of the molecules. We use the pre-calculated grid of slab models ($T,N$) together with a range of emitting radii $R_{\rm e}$ and calculate the emergent slab model spectra $F^{\rm slab}$. The goodness of each slab model for fitting the observations $F^{\rm obs}$ is evaluated using the $\chi^2$ framework
\begin{equation}
\chi^2 = \frac{1}{n} \Sigma_{i=1, n} \frac{(F_i^{\rm obs}-F_i^{\rm slab})}{\sigma^2} \,\,\, .
\end{equation}
Here we assume a constant flux error across all chosen wavelengths windows $\sigma$.

\subsection{Test against 2D thermo-chemical models}

\begin{figure}[t]
\centering
  \includegraphics[width=8cm]{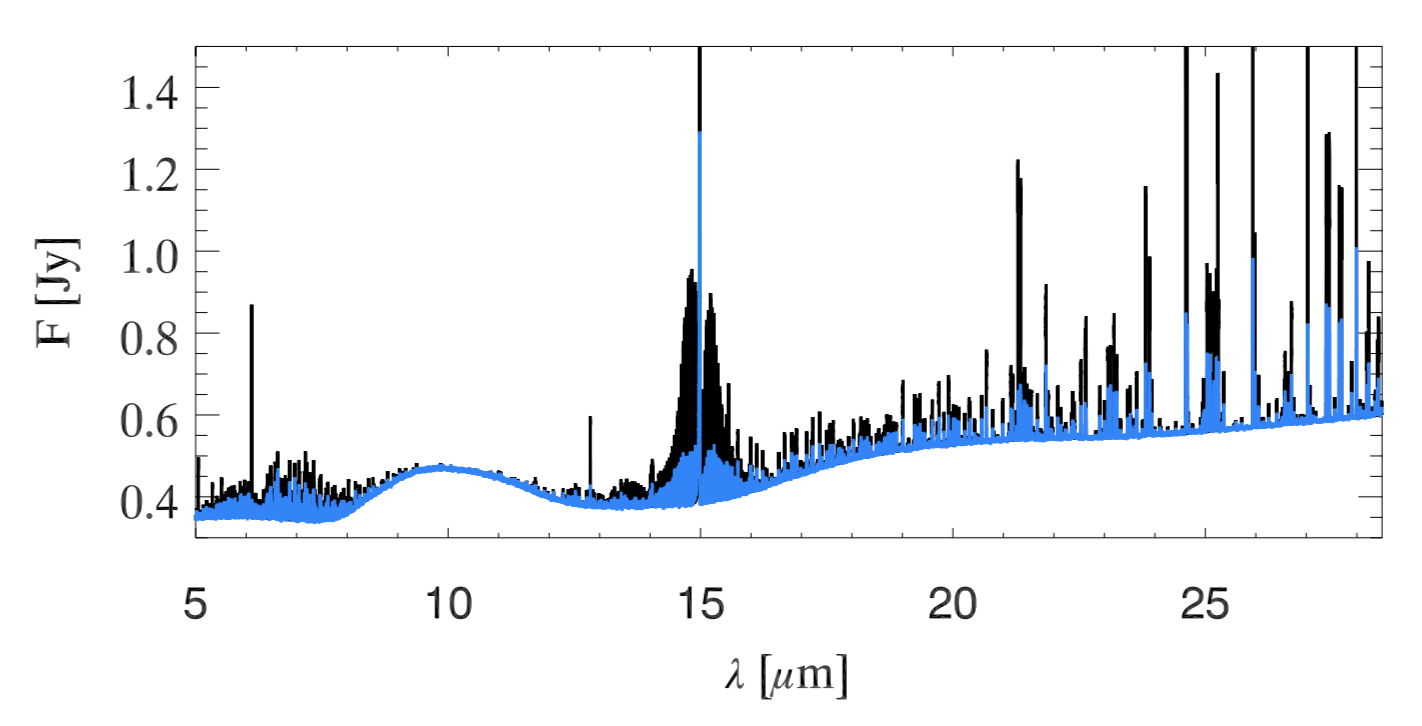}
  \caption{Synthetic mid-IR spectrum for AA\,Tau based on the DIANA model \cite{Woitke2019} using a gas-to-dust ratio of 1000. The black line is the high resolution synthetic model spectrum, the blue one is convolved to $R\!=\!3000$ with noise added (${\rm SNR}\!=\!300$).}
  \label{fig:synth-spectra}
\end{figure}

The above described slab models have been widely used in the analysis of mid-IR spectra. However, it is not clear how the quantities ($T, N, R_{\rm e}$) retrieved relate to the actual properties of the gas in the disk surface. To make a start, we decided to apply the same retrieval technique on a synthetic spectrum generated using a 2D thermo-chemical disk model. For that purpose, we use the codes ProDiMo \cite{Woitke2016, Kamp2017} and FLiTs \cite{Woitke2018}. We take a prominent source AA\,Tau, which has a 2D RT disk model fit to its large multi-wavelengths photometry from the DIANA project \cite{Woitke2019, Dionatos2019}. We calculate the disk chemistry and gas heating/cooling self-consistently using a gas-to-dust ratio of 1000; this is required to ensure that line fluxes are of the same order as those observed by Spitzer \cite{Meijerink2009, Woitke2018}. We use the selection of molecules and molecular data from \cite{Woitke2018}, even though that means using a slightly outdated HITRAN version (2009 instead of 2020). We produce synthetic mid-IR spectra from 5-28.5\,$\mu$m at a nominal spectral resolution of 100\,000 using FLiTs. The spectra are then subsequently convolved to the lower resolution of MIRI ($R=3000$), oversampled at a factor two and we add noise with a typical S/N ratio of 300 (Fig.~\ref{fig:synth-spectra}). We produce these spectra using either all selected molecules simultaneously, or just a single molecule at a time. This is to ensure that we can approach the understanding by stepwise increasing the complexity. Also, the line separation within the ro-vibrational band of CO$_2$ is large enough so that opacity overlap of individual lines is negligible. In the following, we will show the first results using pure CO$_2$ spectra. 

\begin{figure}[t]
\centering
  \includegraphics[width=8cm]{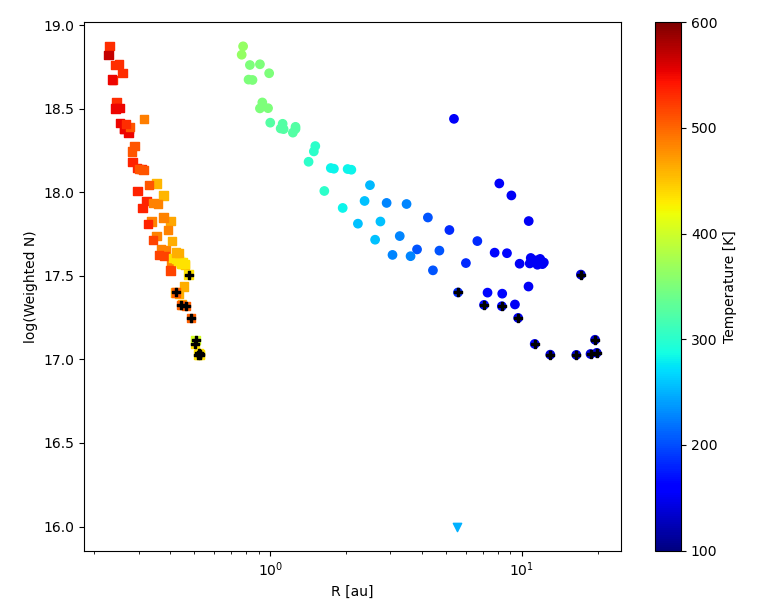}
  \caption{Metadata (minimum, maximum gas temperatures, flux weighted column densities, inner/outer radius of the radial emitting region) for a representative sample of strong CO$_2$ lines extracted from the AA\,Tau thermo-chemical disk model. Black crosses indicate the strongest Q-branch lines. The triangle shows the retrieval result.}
  \label{fig:slab-model-retrieval}
\end{figure}

Figure~\ref{fig:slab-model-retrieval} shows the first results from analysing the emitting conditions in the AA\,Tau thermo-chemical model and comparing them to the results of the slab retrieval. We analyse here only the strongest CO$_2$ lines from the model, but ensure that we have a coverage of P-, Q- and R-branch lines. For this representative sample of lines, we estimate for each line its emitting region by using the 15 and 85\% radial and vertical contribution to the total line flux. We delimit the region at the lower end by the local dust continuum optical depth 1. Over this emitting region, we extract certain metadata for comparison to the retrieval results: the minimum and maximum gas temperatures, flux weighted column densities, the inner/outer radius of the radial emitting region, mass averaged densities of potential collision partners such as electrons, H, H$_2$. It is interesting to note that the emitting region for the different individual CO$_2$ lines in the band span quite a range of radial emitting areas, with the strongest Q-branch lines originating the furthest out and thus also showing the lowest temperatures. It is important to note that the retrieval method is very biased towards the Q-branch when applying the $\chi^2$ fitting approach outlined above. Even more puzzling is then that column density found in the retrieval is an order of magnitude lower than what the thermo-chemical models show. This is only a starting point and more in-depth work is underway, including the effects of opacity overlap within molecular bands and also more elaborate fitting procedures/codes such as CLicK \cite{Liu2019}.

The main warning to take away so far is to not ratio column densities from slab model retrieval to obtain abundances unless it is clear that the two molecules are co-spatial.

\subsection{Degeneracies}

The higher spectral resolution of JWST MIRI/MRS compared to Spitzer/IRS allows us to spectrally resolve the shape of the molecular Q-branches. This is a key element to break the degeneracies between column densities and temperatures in slab models. In several cases \cite{Grant2022, Tabone2022}, the JWST data is now better fit with a high optical depth (high column density) slab model. For optically thin emission large degeneracies remain between the column density and the emitting area.

\subsection{Need for additional molecular data}

Especially the disks around BDs are very rich in hydrocarbons. Only limited data is here available via the HITRAN and GEISA databases. We lack especially the $^{13}$C isotopologues of hydrocarbons, the hot bands of several hydrocarbons (e.g.\ benzene), but also more hydrocarbons in general. The GEISA database lacks the Einstein A coefficients, which are crucial for proper modelling of emission spectra from slabs/disks. We also need fits for partition functions as function of $T, P$, covering the typical parameter space of inner disks ($T\!=\!100\ldots5000$~K, $P\!=\!10^{-6}-1$~mbar).

\section{MINDS results}

We focus here on two types of disks, around T\,Tauri stars and around very low-mass stars (VLMS) or brown dwarfs (BDs). 
Their chemistry has already been found to be vastly different with Spitzer. The VLMS have higher C$_2$H$_2$/HCN and HCN/H$_2$O column density ratios than the T\,Tauri disks \citep{Pascucci2009,Pascucci2013}.
Clearly, the higher spectral resolution and sensitivity of JWST MIRI/MRS allow us to study these differences in much more detail, hopefully unraveling the cause of this profound dichotomy.

\subsection{Disks around T\,Tauri stars}

\begin{figure*}[t]
\centering
 \includegraphics[width=14cm]{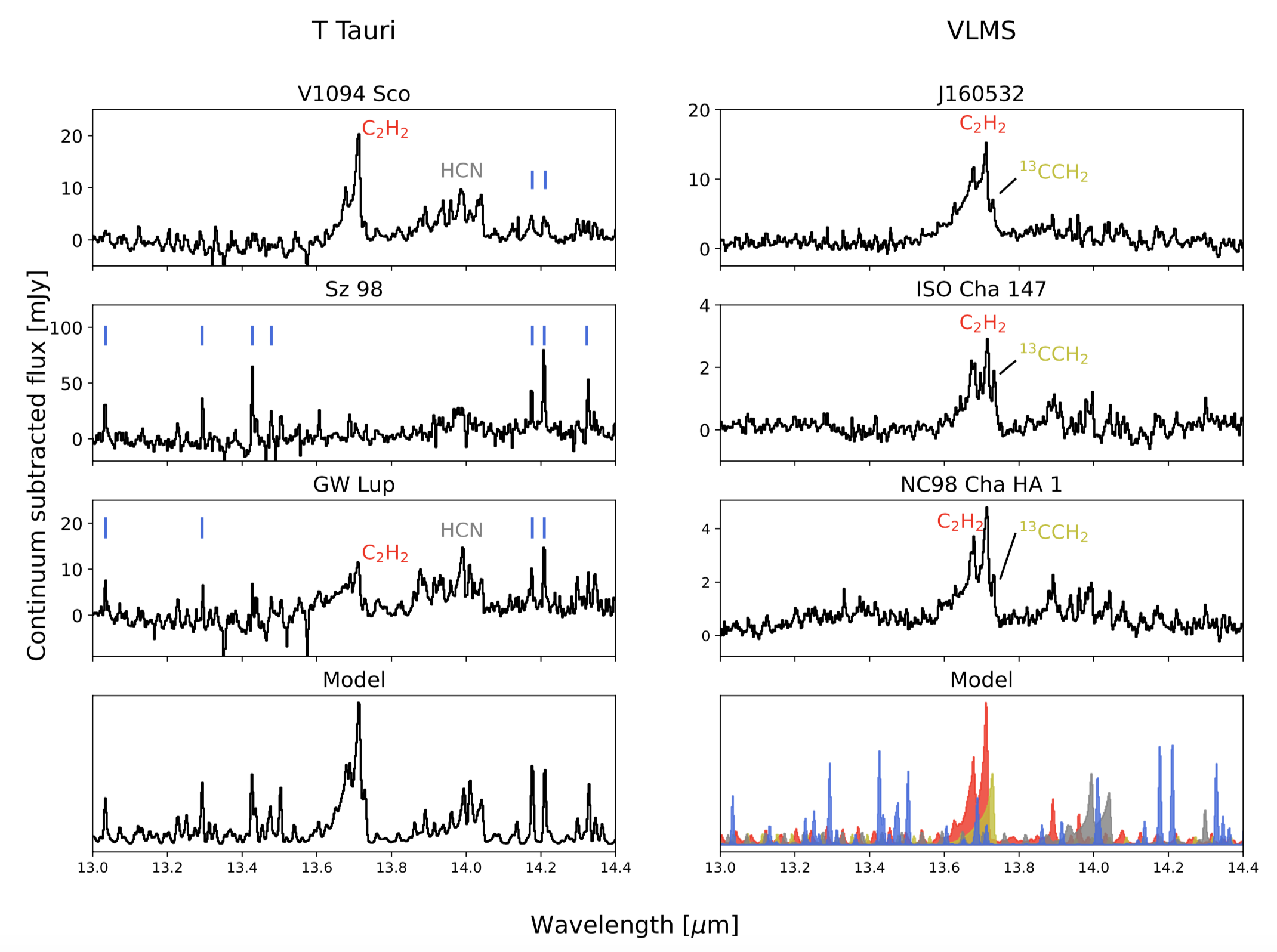}
 \caption{Left: MIRI/MRS spectra of the T\,Tauri stars V1094\,Sco, Sz\,98 and GW\,Lup. Right: MIRI/MRS spectra of the VLMS and BDs J160532, ISO\,Cha\,147 and NC98\,Cha\,H$\alpha$1. The lowest panel on each side shows LTE slab models ($T=500$~K, $N=10^{16}$~cm$^{-2}$, arbitrary emitting area to match observed spectra) of the main molecules detected in this wavelength range, C$_2$H$_2$ (red), $^{13}$CCH$_2$ (gold), HCN (grey) and water (blue --- vertical tickmarks). The lower left panel shows them co-added, the lower right panel individually with separate colors (see above).}
 \label{fig:MIRI-spectra}
\end{figure*}

We show here a selected wavelength range centered on the two molecules C$_2$H$_2$ and HCN, including also a few stronger water lines (13 - 14.3~$\mu$m). Figure~\ref{fig:MIRI-spectra} (left panel) shows the diversity in relative feature strength among three of our T\,Tauri sources: V1094\,Sco (K6, $d=154$~pc, $L_\ast=1.15$~L$_\odot$ \cite{Sanchis2020}), Sz\,98 (K7, $d=156.22$~pc, $L_\ast=1.53$~L$_\odot$ \cite{vanTerwisga2019}), and GW\,Lup (M1.5, $d=155$~pc, $L_\ast=0.33$~L$_\odot$ \cite{Alcala2017,Andrews2018}).

GW\,Lup has been previously found to be a CO$_2$ rich source \cite{Pontoppidan2010} and \cite{Grant2022} report now also the first detection of $^{13}$CO$_2$ and water in this source. The other two sources, Sz\,98 and V1094\,Sco, have only low-resolution Spitzer spectra taken and thus no previous molecule identifications in the mid-IR. Sz\,98 is the strongest water source in this sample of three. Also, all three molecules, C$_2$H$_2$, water and HCN are clearly detected in V1094\,Sco; however, the molecular flux ratio C$_2$H$_2$ to HCN is higher than in GW\,Lup, and the water lines are weaker.

Thus, the MIRI/MRS spectra show a pronounced diversity in feature ratio between C$_2$H$_2$ and HCN, but also in the relative strength of water line fluxes (indicated in blue in Fig.~\ref{fig:MIRI-spectra}). The slab models in the lowest left panel of Fig.~\ref{fig:MIRI-spectra} are co-added to illustrate the richness in substructure in the HCN ro-vibrational wavelength range; with the higher spectra resolution of MIRI/MRS versus Spitzer, we can clearly resolve the structure and see the individual contribution of also C$_2$H$_2$ and water lines in there. The question whether the diversity seen in the three T\,Tauri spectra is related to disk physical structure (gas+dust gaps, flaring, heating, radial transport), different dust properties (settling, sizes, gaps), or some other process remains open and will be discussed in Sect.~\ref{Sec:discussion}.

\subsection{Disks around VLMS and brown dwarfs}

Similar to the T\,Tauri stars, we show here also three disks around VLMS/BDs: 2MASS-J16053215-1933159 (J160532, M4.75, $d=152$~pc, $L_\ast=0.04$~L$_\odot$ \cite{Pascucci2013,Luhman2018}), ISO\,Cha\,147 (M5.75, $d=200$~pc, $L_\ast=0.01$~L$_\odot$ \cite{Pascucci2009,Manara2019,Manara2017}), NC98\,Cha\,H$\alpha$\,1 (M7.5, $d=200$~pc, $L_\ast=0.015$~L$_\odot$ \cite{Manara2017,Luhman2007}).

From Fig.~\ref{fig:MIRI-spectra}, we see that the VLMS/BD disks show structurally very strong C$_2$H$_2$ emission as already found from Spitzer spectra \cite{Pascucci2009,Pascucci2013}. However, with the higher spectral resolution of MIRI/MRS, we now also detect in all three disks the isotopologue $^{13}$CCH$_2$ \cite{Tabone2022}. This extra Q-branch on the red shoulder of the main isotopologue (see lowest right panel in Fig.~\ref{fig:MIRI-spectra}, red and gold) is very pronounced in these disks compared to the T\,Tauri disks. Hence, they likely have much larger acetylene column densities or an unusually high isotope ratio $^{13}$C/$^{12}$C. In fact, the LTE slab model fit to the MIRI spectrum of J160532 indicates an acetylene column density of $N=3.2\,10^{20}$~cm$^{-2}$, making the emission highly optically thick and leading to a quasi-continuum \cite{Tabone2022}. Due to the overlap of these isotopologues, the slab model fitting should ideally be done for both simultaneously and likely opacity overlap not only plays a role for within a molecule, but possibly also among isotopologues. More importantly, we are currently limited by the availability of molecular data for higher excited bands of the rare isotopologue.

In the disk around J160532, we also detected two new hydrocarbons, C$_4$H$_2$ and C$_6$H$_6$ ($N\,\sim\,10^{17}$~cm$^{-2}$); there is also a tentative detection of CH$_4$. The emitting radius (circular emitting area $\pi R^2$) of these hydrocarbon molecules is very small within $\sim0.07$~au. If this were to correspond to the physical radius in the disk from which the molecules emit, it would lie within the snowline expected of a typical BD disk \cite{walsh2015,Greenwood2017}.

What is also striking is the absence of strong warm water emission in these observed disk spectra. The features at $14.2$~$\mu$m, which are clearly attributed to water in the T\,Tauri disks, originate from C$_2$H$_2$ in these objects (see lowest right panel in Fig.~\ref{fig:MIRI-spectra} for the overlap of those species).

\section{Discussion}
\label{Sec:discussion}

\subsection{Diversity in T\,Tauri disks: Observations versus models}

The diversity in T\,Tauri disk spectra is very intriguing. Many individual parameter studies have been carried out earlier and we summarize here the key findings in the light of how they affect mid-IR spectra specifically.

The gas-to-dust mass ratio in the disk surface directly impacts the strength of the molecular emission lines \cite{Meijerink2009, Antonellini2015}. For disks with a nominal gas-to-dust mass ratio of 100, the dust continuum limits the molecular line emitting region vertically; hence many disk models adopted a value of 1000 to match observed line fluxes (see also Fig.~\ref{fig:synth-spectra}). The more the disk surface flares, the more radiation it intercepts and the warmer the gas becomes. This also boosts the strength of the molecular emission lines, radially extending the emitting area of mid-IR molecular lines \cite{Antonellini2015, Greenwood2019b}. The dust properties in the inner disk surface are strongly affected by settling, radial drift, growth, all of which can lower the local dust opacity. Simple parameter studies in radiation thermo-chemical models have shown that molecular lines become stronger if the local dust opacity is lowered \cite{Antonellini2015}. 

Connecting the basic understanding from those studies to disk evolutionary processes is a next step. A simplified dust evolution following the two population approach \cite{Birnstiel2012} naturally lowers the dust opacity in the inner disk and together with settling this produces high gas-to-dust mass ratios in the disk surface \cite{Greenwood2019a}; all of these effects lead to a strengthening of mid-IR spectra in the presence of dust evolution. Ice transport along with radially drifting dust grains is shown to be very efficient and strongly affects the mid-IR spectra \cite{Bosman2018}. Coupling dust growth (pebble formation), dynamics and chemistry leads to an increasing complexity in predicting inner disk composition \cite{Krijt2020, Eistrup2022}; the detailed composition will also depend on the presence of disk substructure (pressure bumps that keep icy pebbles from drifting \cite{Pinilla2012}), the cold finger effect (freezing out of ices enhancing growth and settling rates \cite{Meijerink2009, Krijt2016}), and details of ice condensation/diffusion/sublimation processes \cite{Cuppen2017, Powell2022}. 

Apart from the above advances in our understanding of how dust opacity, chemistry and disk evolution are intertwined, there are also more intricate effects directly related to molecules and line radiative transfer. The self-shielding of molecules and/or mutual shielding is a key example of that. Recent work \cite{Bethell2009, Bosman2022a, Bosman2022b} shows that this has a profound effect on the warm inner disk chemistry since water self-shielding prevents the O to be unlocked as OH and proceed to form CO$_2$. This affects the vertical location of the abundant warm reservoirs of water and CO$_2$ and hence also the relative strength of the CO$_2$-to-H$_2$O emission features.

\begin{figure*}[t]
\centering
 \includegraphics[width=8cm]{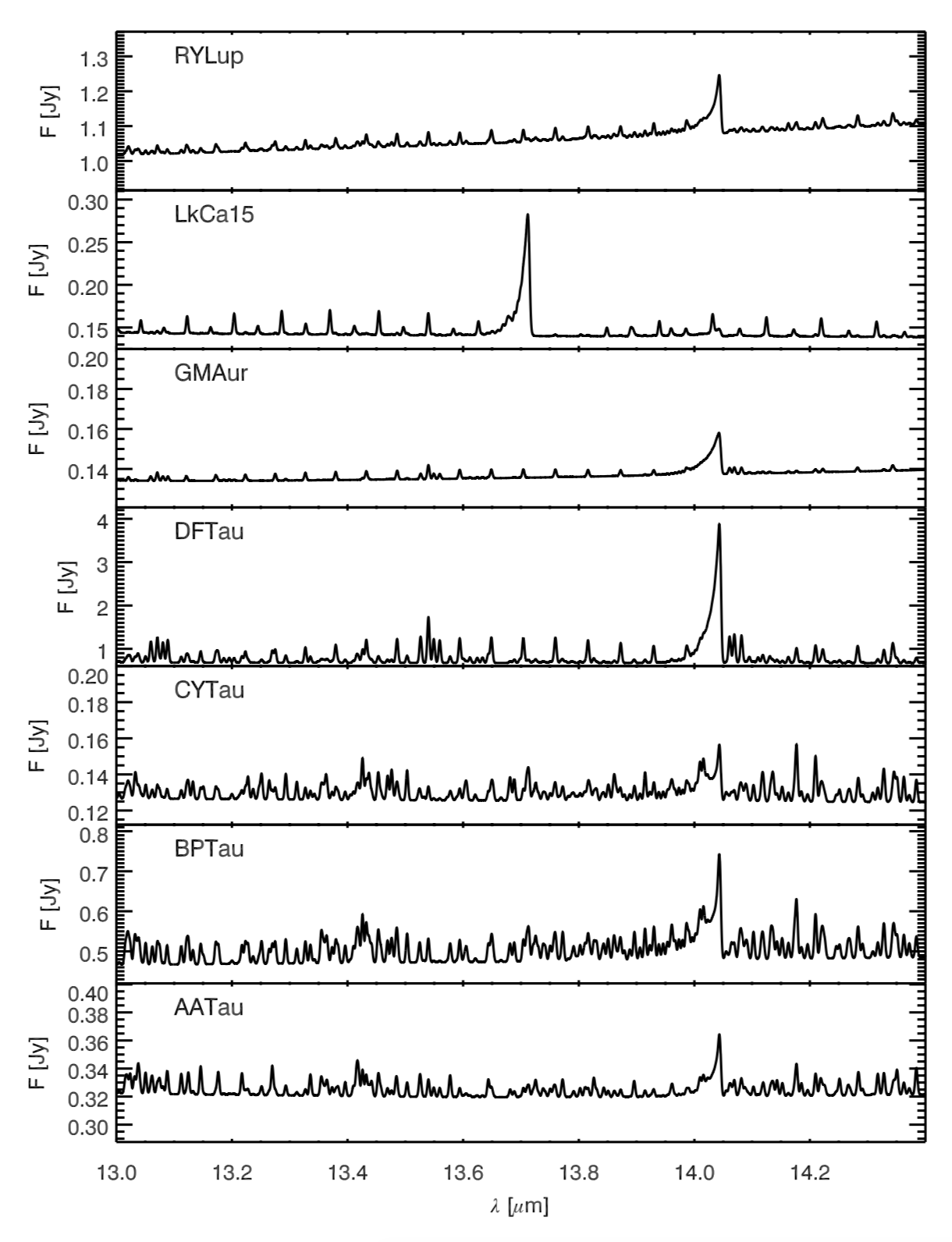}
 \includegraphics[width=8.3cm]{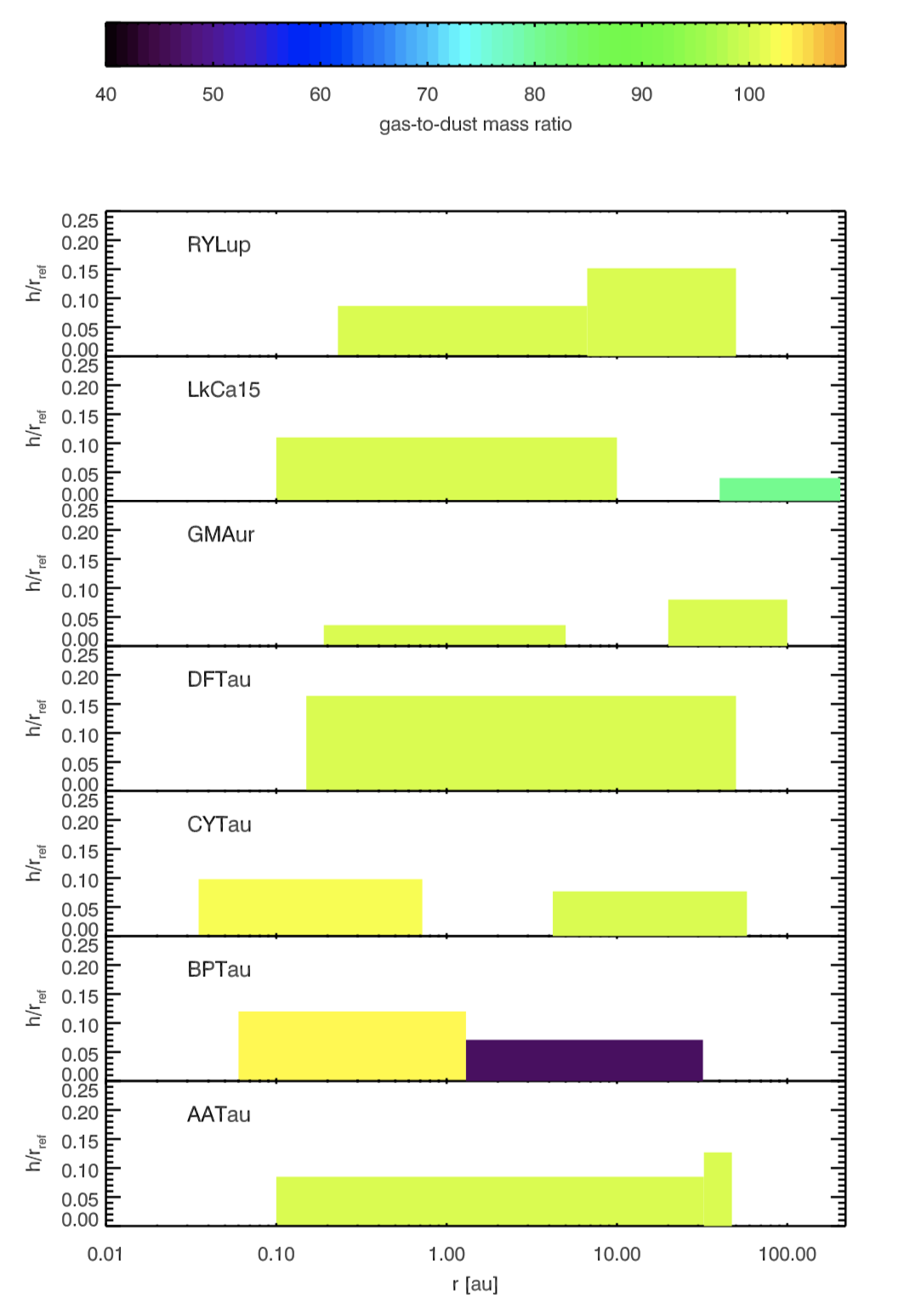}
 \caption{Left: Mid-infrared spectra predicted from a selected set of T\,Tauri DIANA disk models. Right: Schematic structure for those selected DIANA models, showing the radial disk structure, the dimensionless scale height (scale height divided by the reference radius for the inner/outer zone respectively) and the gas-to-dust mass ratio in color.}
 \label{fig:DIANA-spectra-structure}
\end{figure*}

Instead of parametrized standard disk models that are often used in building up understanding of the coupling of dynamics, physics, chemistry, and radiative transfer, we can also approach the interpretation from the perspective of observational diversity in disk structure. The DIANA project \cite{Woitke2016, Kamp2017, Woitke2019, Dionatos2019} has determined the physical parameters of 14 disks using multi-wavelengths observational data and a coherent radiation thermo-chemical disk modeling approach including dust and gas. Figure~\ref{fig:DIANA-spectra-structure} shows mid-IR spectra computed for a subset of disk models from this project. We took the models and calculated mid-IR spectra using the same selection for the molecules as used in \cite{Woitke2018}. We show here only the results using the simplified escape probability treatment for the line fluxes. This sample covers spectral types K5 to M1, so a range similar to the presented observed three sources. The fitting approach allowed the use of two zones that are either directly connected or separated by a gap (for details of the modeling approach and the fitting strategy, see \cite{Woitke2016, Woitke2019}). We show here the same wavelength range as for the MIRI/MRS observations and also convolved the simulated spectra with a resolution of $R=3000$. It is very intriguing to see that also these models cover a diversity of molecular feature ratios. Note however that these models were not tailored to fit any specific mid-IR line fluxes. The only point to be made is that the detailed inner disk structure, i.e.\ location of gaps, variation in scale height, gas-to-dust mass ratio affect the various molecular features in very different ways.

\subsection{VLMS and brown dwarf disks}
\label{Sec:Discussion-VLMSandBDs}

The fact that these stars have much lower luminosity compared to their T\,Tauri counterparts is to some extent offset by the dust sublimation radius being closer to the star as well. So, to some extent, the disk surface should be self-similar in temperatures and chemistry, except that the warm region is spatially less extended and thus emitting areas should be much smaller compared to T\,Tauri disks \cite{Walsh2015, Greenwood2017}. There seems to be also no indication that the turbulence works differently across this mass range \cite{Mulders2012}. Herschel observations found that the disks around BDs may have a lower disk-to-star mass fraction than T\,Tauris \cite{Harvey2012}. Various studies also show that the dynamical processes occurring in disks around VLMS and BDs can be different from their more massive counterparts. For example, dust evolution has been shown to proceed much faster in these lower mass objects \cite{Pinilla2017}.

The spectra of the disks around VLMS and BDs suggest that we are witnessing the carbon chemistry that proceeds in environments with a C/O ratio larger than one. The formation pathways for hydrocarbons in disks, especially C$_2$H$_2$ have been recently revisited (Kanwar submitted). The surface reservoir of this molecule is driven by chemical pathways that involve H, H$_2$ and C to be abundant. C unlocking occurs via UV or X-ray dissociation of CO. Then there are neutral-neutral and ion-molecule pathways to drive the formation of C$_2$H$_2$. A high C/O ratio should allow for a much more radially and vertically extended hydrocarbon reservoir. The impact of this element ratio has been explored only for typical T\,Tauri disks \cite{Najita2011, Pascucci2013, Woitke2018, Anderson2021} and shown to affect water, CO$_2$, HCN and C$_2$H$_2$ fluxes; the latter two increasing as C/O is pushed to values larger than 1. HCN increases gradually with an increasing C/O ratio, while C$_2$H$_2$ remains constant and starts increasing for C/O$\,>\,1$.

The extremely high column densities of C$_2$H$_2$ observed in J160532 could be directly linked to the absence of silicate features in that disk \cite{Tabone2022}. If the absence of a clear silicate emission feature is due to the average grain size in the line emitting region being larger than a few $\mu$m, then this implies a low opacity, and hence allows us to 'see' deep into the disk, and probe the full gas column almost down to the midplane. This difference in dust content in disks around VLMS and BDs could be linked to the faster dust radial migration discussed above.

\subsection{Retrieval strategies}

The limitations of using 0D slab models for retrieval are quite clear: single molecule, constant temperature, and density. From radiation thermo-chemical disk models, we know that the molecular line emitting regions have radial and vertical density and temperature gradients. Also, the different molecules are not necessarily co-spatial. Hence, an approach that allows for a radial temperature gradient like CLicK \cite{Liu2019} is already a large improvement. The strong C$_2$H$_2$ band in the BD J160532 that require a combination of optically thick and thin slab models to fit could indicate that we are in fact seeing that the emission is not confined to a narrow region in the disk (as assumed in a 0D slab model).

Even if sticking with 0D slab models, instead of a sequential fitting from the molecules with strongest flux contribution down to the ones with least flux contribution, simultaneous fitting of all molecules within an MCMC approach should be developed. The largest issue with slab models pertaining to single molecules could be opacity overlap for isotopologues.

A full radiation thermo-chemical disk modeling is beyond reach for retrieval \cite{Woitke2019}. However, disk models fitted to a large range of observables of individual sources provide a great laboratory to enable deeper physical understanding and context even to the 0D slab model results.

\section{Outlook}

JWST MIRI/MRS has tremendous potential to revolutionize our understanding of the chemistry in the inner 10~au of planet-forming disks. From the results shown above, we are left with some key questions: Is the dichotomy between T\,Tauri and BD disk chemical composition a real dichotomy or is there a continuous spectrum? If there is a dichotomy, where is the break in SpType? What causes the difference in chemistry? Why would the C/O ratio be different in disks around VLMS and brown dwarfs? Are the hydrocarbons that we detect in disks around VLMS and BDs efficiently formed (combustion chemistry) from bottom up or do we witness the breaking down of large PAHs/carbonaceous dust?

We urgently need complete molecular data (including Einstein A coefficients and hot bands) for hydrocarbons, but also isotopologues.

\section*{Author Contributions}

Investigation: Arabhavi, Argyriou, Bettoni, Christiaens, Gasman, Grant, Guadarrama, Jang, Kanwar, Morales-Calderon, Pawellek, Perotti, Rodgers-Lee, Samland, Schreiber, Scheithauer, Schwarz, Tabone, Temmink, Vlasblom;
Funding acquisition: Henning, Kamp, van Dishoeck, G\"{u}del; 
Supervision: Abergel, Barrado, Bouwman, Boccaletti, Caratti o Caratti, van Dishoeck, Geers, Glauser, G\"{u}del, Henning, Lagage, Lahuis, M\"{u}ller, Nehm\'{e}, Olofsson, Pantin, Ray, Vandenbussche, Waelkens, Waters, Wright;
Software: Arabhavi, Grant, Kamp, Schwarz, Tabone, Waters; 
Writing – original draft: Kamp, Christiaens; Writing – review \& editing: Henning, all authors.

\section*{Conflicts of interest}

There are no conflicts to declare.

\section*{Acknowledgements}

The following National and International Funding Agencies funded and supported the MIRI development: NASA; ESA; Belgian Science Policy Office (BELSPO); Centre Nationale d’Etudes Spatiales (CNES); Danish National Space Centre; Deutsches Zentrum fur Luftund Raumfahrt (DLR); Enterprise Ireland; Ministerio De Econom\'ia y Competividad; Netherlands Research School for Astronomy (NOVA); Netherlands Organisation for Scientific Research (NWO); Science and Technology Facilities Council; Swiss Space Office; Swedish National Space Agency; and UK Space Agency.
I.K.\ acknowledges support from the H2020-MSCA-ITN-2019 grant no.\ 860470 (CHAMELEON). 
I.K., A.M.A., and E.v.D.\ acknowledge support from grant TOP-1614.001.751 from the Dutch Research Council (NWO).
I.K.\ and J.K.\ acknowledge funding from H2020-MSCA-ITN-2019, grant no. 860470 (CHAMELEON).
A.C.G.\ has been supported by PRIN-INAF MAIN-STREAM 2017 “Protoplanetary disks seen through the eyes of new-generation instruments” and from PRIN-INAF 2019 “Spectroscopically tracing the disk dispersal evolution (STRADE)”.
G.B.\ thanks the Deutsche Forschungsgemeinschaft (DFG) - grant 138 325594231, FOR 2634/2.
E.v.D.\ acknowledges support from the ERC grant 101019751 MOLDISK and the Danish National Research Foundation through the Center of Excellence ``InterCat'' (DNRF150). 
D.G.\ would like to thank the Research Foundation Flanders for co-financing the present research (grant number V435622N).
T.H.\ and K.S.\ acknowledge support from the European Research Council under the Horizon 2020 Framework Program via the ERC Advanced Grant Origins 83 24 28. 
B.T.\ is a Laureate of the Paris Region fellowship program, which is supported by the Ile-de-France Region and has received funding under the Horizon 2020 innovation framework program and Marie Sklodowska-Curie grant agreement No.\ 945298.
O.A.\ and V.C.\ acknowledge funding from the Belgian F.R.S.-FNRS.
I.A.\ and D.G.\ thank the European Space Agency (ESA) and the Belgian Federal Science Policy Office (BELSPO) for their support in the framework of the PRODEX Programme.
T.P.R.\ acknowledges support from ERC grant 743029 EASY.
D.R.L.\ acknowledges support from Science Foundation Ireland (grant number 21/PATH-S/9339).
D.B.\  and M.M.C.\ have been funded by Spanish MCIN/AEI/10.13039/501100011033 grants PID2019-107061GB-C61 and No. MDM-2017-0737 

\footnotetext{\textit{$^{c}$~Universit\'e Paris-Saclay, CNRS, Institut d’Astrophysique Spatiale, 91405, Orsay, France. }}

\footnotetext{\textit{$^{d}$~STAR Institute, Universit\'e de Li\`ege, All\'ee du Six Ao\^ut 19c, 4000 Li\`ege, Belgium. }}

\footnotetext{\textit{$^{e}$~Institute of Astronomy, KU Leuven, Celestijnenlaan 200D, 3001 Leuven, Belgium. }}

\footnotetext{\textit{$^{f}$~Centro de Astrobiolog\'ia (CAB), CSIC-INTA, ESAC Campus, Camino Bajo del Castillo s/n, 28692 Villanueva de la Ca\~nada,
Madrid, Spain. }}

\footnotetext{\textit{$^{g}$~Max-Planck Institut f\"{u}r Extraterrestrische Physik (MPE), Giessenbachstr.\ 1, 85748, Garching, Germany. }}

\footnotetext{\textit{$^{h}$~LESIA, Observatoire de Paris, Universit\'e PSL, CNRS, Sorbonne Universit\'e, Universit\'e de Paris, 5 place Jules Janssen, 92195 Meudon, France. }}

\footnotetext{\textit{$^{i}$~INAF – Osservatorio Astronomico di Capodimonte, Salita Moiariello 16, 80131 Napoli, Italy. }}

\footnotetext{\textit{$^{j}$~Dublin Institute for Advanced Studies, 31 Fitzwilliam Place, D02 XF86 Dublin, Ireland. }}

\footnotetext{\textit{$^{k}$~Leiden Observatory, Leiden University, 2300 RA Leiden, the Netherlands. }}

\footnotetext{\textit{$^{l}$~UK Astronomy Technology Centre, Royal Observatory Edinburgh, Blackford Hill, Edinburgh EH9 3HJ, UK. }}

\footnotetext{\textit{$^{m}$~ETH Z\"urich, Institute for Particle Physics and Astrophysics, Wolfgang-Pauli-Str. 27, 8093 Z\"urich, Switzerland. }}

\footnotetext{\textit{$^{n}$~Dept. of Astrophysics, University of Vienna, T\"urkenschanzstr 17, A-1180 Vienna, Austria. }}

\footnotetext{\textit{$^{o}$~Department of Astrophysics/IMAPP, Radboud University, PO Box 9010, 6500 GL Nijmegen, The Netherlands. }}

\footnotetext{\textit{$^{p}$~Space Research Institute, Austrian Academy of Sciences, Schmiedlstr. 6, A-8042, Graz, Austria. }}

\footnotetext{\textit{$^{q}$~Universit\'e Paris-Saclay, Universit\'e de Paris, CEA, CNRS, AIM, F-91191 Gif-sur-Yvette, France. }}

\footnotetext{\textit{$^{r}$~SRON Netherlands Institute for Space Research, PO Box 800, 9700 AV, Groningen, The Netherlands. }}

\footnotetext{\textit{$^{s}$~CEA/DSM/Irfu/Service d'Astrophysique - Laboratoire AIM. }}

\footnotetext{\textit{$^{t}$~Department of Astronomy, Stockholm University, AlbaNova University Center, 10691 Stockholm, Sweden. }}

\footnotetext{\textit{$^{u}$~IRFU/DAp D\'epartement D’Astrophysique CE Saclay, Gif-sur-Yvette, France. }}

\footnotetext{\textit{$^{v}$~SRON Netherlands Institute for Space Research, Niels Bohrweg 4, NL-2333 CA Leiden, the Netherlands. }}

\footnotetext{\textit{$^{w}$~Konkoly Observatory, Research Centre for Astronomy and Earth Sciences, E\"otv\"os Lor\'and Research Network (ELKH), Konkoly-Thege Mikl\'os \'ut 15-17, H-1121 Budapest, Hungary}}



\balance


\bibliography{bibliography} 
\bibliographystyle{bibliography} 

\end{document}